\documentclass[preprint,12pt]{elsarticle}
\usepackage{amssymb,epsfig,graphicx,multirow,bm,dcolumn}
\journal{JALCOM}
\begin{document}
\begin{frontmatter}

\title{Importance of site occupancy and absence of strain glassy phase in Ni$_{2-x}$Fe$_{x}$Mn$_{1.5}$In$_{0.5}$}
\author[gu]{R. Nevgi}
\author[jnc]{Gangadhar Das}
\author[dui]{M. Acet}
\author[gu]{K. R. Priolkar \corref{krp}}\ead{krp@unigoa.ac.in}
\cortext[krp]{Corresponding author}
\address[gu]{Department of Physics, Goa University, Taleigao Plateau, Goa 403206 India}
\address[jnc]{Chemistry and Physics of Materials Unit, Jawaharlal Nehru Centre for Advanced Scientific Research, Jakkur, Bengaluru, 560064 India}
\address[dui]{Experimentalphysik, University of Duisburg-Essen, 47048 Duisburg, Germany}

\begin{abstract}
Martensitic transition temperature steadily decreases in Ni$_{2-x}$Fe$_{x}$Mn$_{1.5}$In$_{0.5}$ and is completely suppressed at $x$ = 0.2. Despite suppression of martensitic transition, Ni$_{1.8}$Fe$_{0.2}$Mn$_{1.5}$In$_{0.5}$ does not display the expected strain glassy phase. Instead, a ground state with dominant ferromagnetic interactions is observed. A study of structural and magnetic properties of $x$ = 0.2 reveal that the alloy consists of a major Fe rich cubic phase and a minor Fe deficient monoclinic phase favoring a ferromagnetic ground state. This is exactly opposite of that observed in Ni$_2$Mn$_{1-y}$Fe$_{y}$In$_{0.5}$ wherein a strain glassy phase is observed for $y$ = 0.1. The change in site symmetry of Fe when doped for Ni in contrast to Mn in the Heusler composition seems to support the growth of the ferromagnetic phase.
\end{abstract}

\begin{keyword}
Strain glass, Ni$_{2-x}$Fe$_{x}$Mn$_{1.5}$In$_{0.5}$, Ferromagnet
\end{keyword}
\end{frontmatter}

\section{Introduction}

Strain glass phase occurs in an impurity doped ferroelastic alloy when the dopant concentration exceeds a certain limit. This phase is analogous to a spin glass or a cluster glass phase in magnetic systems \cite{Zhou200995, Zhou201058}. Strain glass phase is a frozen disordered ferroelastic phase with a short-range strain order formed in a material undergoing a first-order transformation \cite{Sarkar200595, Lloveras2008100, Vasseur201082}. Martensitic to strain glass transition is considered to be a natural consequence of impurity doping and has been shown to occur in many different impurity doped martensitic alloys like NiTi, TiPd, Au-Cu-Al, FeNiMnC, etc. \cite{Wang201058, Zhou201058, Zhou200995, MA201874}.

A material undergoing martensitic transformation exhibits a structural change from a high symmetry austenitic structure (usually cubic) to a lower symmetry martensitic structure. On the other hand, the strain glassy phase consists of nano-sized domains with frozen elastic strain vector, and as such, the crystal structure of such material remains invariant across the transition temperature $T_g$ \cite{Ren2012}. The important characteristic of a strain glassy phase, however, is the behavior of the dip in storage modulus and peak in the imaginary part of dynamical mechanical properties (loss or $\tan\delta$) in accordance with the Vogel-Fulcher law \cite{Wang200867, Ren201090}. Another characteristic that distinguishes a strain glassy phase from pre-martensitic tweed is the existence of ergodicity breaking as evidenced in zero field cooled (ZFC) and field cooled (FC) experiments.

Recently, an unusual strain glassy phase was reported in Ni$_2$Mn$_{1.4}$Fe$_{0.1}$In$_{0.5}$ with a $T_g$ = 350K \cite{Nevgi2018112}. The alloy, though has an incommensurate 7M modulated martensitic structure at room temperature, exhibits other glassy characteristics like the frequency dependent behavior of storage modulus and loss obeying Vogel Fulcher law and ergodicity breaking between ZFC and FC measurements just above the transition temperature. The existence of modulated structure was explained due to the presence of strain domains large enough to present signatures of long-range martensitic order in diffraction measurements but remain non interacting with each other due to the presence of an minor impurity phase rich in Fe \cite{Nevgi2018112}.

While strain glassy phase is quite common in binary alloys like NiTi, AuCu, etc., relatively few Heusler alloys have been reported to exhibit strain glass phase. Apart from Ni$_2$Mn$_{1.5-x}$Fe$_{x}$In$_{0.5}$, Co doped Ni-Mn-Ga (Ni$_{55-x}$Co$_x$Mn$_{20}$Ga$_{25}$) is the only other Heusler alloy to exhibit strain glassy phase \cite{Wang201298}. In an X$_2$YZ Heusler alloy, the site symmetry of the X site ($\bar43m$) is different from that of Y and Z sites ($m\bar3m$), and this difference is known to affect structural and magnetic interactions. Therefore, it would be interesting to explore the effect of site occupancy on the occurrence of strain glassy phase in Heusler alloys. Presence of strain glassy phase in Ni$_2$Mn$_{1.5-x}$Fe$_{x}$In$_{0.5}$, wherein Fe is doped for Mn at the Y/Z site is already documented \cite{Nevgi2018112}.  Here, we explore the effect of Fe doping at Ni (X) site in the same martensitic composition of Ni$_2$Mn$_{1.5}$In$_{0.5}$ to realize Ni$_{2-x}$Fe$_x$Mn$_{1.5}$In$_{0.5}$ ($x$ = 0, 0.1, 0.15, 0.2). The main objective is to test the universality of strain glassy phase in Heusler alloys and to understand site dependent preferences of the impurity atoms, if any, in driving such non-ergodic transitions. Through a study of structural, mechanical, transport and magnetic properties of Ni$_{2-x}$Fe$_x$Mn$_{1.5}$In$_{0.5}$, we report absence of strain glassy phase in $x$ = 0.2 alloy composition and attribute it to a phase separation of the alloy into a mixture consisting of majority ferromagnetic austenitic phase that is rich in Fe and a minority martensitic phase that is poor in Fe content. This situation is interesting because such a phase separation is exactly opposite to that occurring in strain glassy Ni$_2$Mn$_{1.4}$Fe$_{0.1}$In$_{0.5}$.

\section{Experimental}

The alloys, Ni$_{2-x}$Fe$_x$Mn$_{1.5}$In$_{0.5}$ ($x$ = 0, 0.1, 0.15, 0.2), were prepared by arc melting in argon atmosphere ensuring stoichiometric proportions of each constituent element. The homogeneity of individual alloy mixture is guaranteed by flipping over the ingot several times during the preparation process. A part of every bead was cut, and remaining was powdered. The cut bead and the powder covered in tantalum foil were encapsulated in an evacuated quartz tube, annealed at 750$^\circ$C for 48 hours and subsequently quenched in ice cold water. The alloys so formed were subjected to scanning electron microscopy with energy dispersive x-ray (SEM-EDX) measurements and the compositions obtained are reported in Table \ref{table:EDX}. In general an error of 2--5\% was noted in the compositions of the alloys.  In order to obtain room temperature structural information, x-ray diffraction (xrd) patterns were recorded on powdered samples using Mo K$_\alpha$ radiation in the angular range of 10$^\circ$ to 70$^\circ$. Synchrotron x-ray diffraction measurement was also achieved on BL-18B at Photon Factory, KEK, Tsukuba, Japan using incident photons of 16 KeV. The diffraction patterns were recorded at 300 K and 500 K to study the structural changes occurring with Fe doping. The martensitic transformation temperatures were confirmed using differential scanning calorimetry (DSC) and four probe resistivity measurements. DSC measurements were performed using Shimadzu DSC-60 on 7-8 mg pieces of each alloy crimped in aluminium pan by heating/cooling at a constant rate of 5$^\circ$C/min. The resistivity measurements were carried out by normal four probe method on rectangular pieces of dimensions 9.7 mm $\times$ 3 mm $\times$ 1 mm. Dynamic mechanical analyzer (DMA) (Q800, TA Instruments) was used to obtain frequency-dependent measurements of AC storage modulus and internal friction (tan$\delta$). Measurements were performed using 3 point bending mode by applying a small AC stress that generated a maximum displacement of 5 $\mu$m at different frequencies in the range of 0.1 Hz to 7 Hz on rectangular pieces of (10 mm $\times$ 5 mm $\times$ 1 mm) dimensions. Temperature-dependent magnetization measurements were carried out using a SQUID magnetometer in the applied magnetic field of 50 Oe in the temperature range of 10 K to 380 K during zero field cooled (ZFC), field cooled cooling (FCC), and field cooled warming (FCW) cycles.

\begin{table}[h]
\caption{Nominal and Actual composition as obtained from SEM-EDX analysis of the four alloys.}
\vspace{0.3cm}
\centering
\begin{tabular}{ccc}
\hline\hline
Sr No. & Nominal composition & Actual composition \\ [0.5ex]
\hline
1 & Ni$_{2}$Mn$_{1.5}$In$_{0.5}$                & Ni$_{2.01}$Mn$_{1.45}$In$_{0.53}$\\
2 & Ni$_{1.9}$Fe$_{0.1}$Mn$_{1.5}$In$_{0.5}$    & Ni$_{1.88}$Fe$_{0.13}$Mn$_{1.48}$In$_{0.51}$\\
3 & Ni$_{1.85}$Fe$_{0.15}$Mn$_{1.5}$In$_{0.5}$  & Ni$_{1.81}$Fe$_{0.18}$Mn$_{1.51}$In$_{0.5}$\\
4 & Ni$_{1.8}$Fe$_{0.2}$Mn$_{1.5}$In$_{0.5}$    & Ni$_{1.75}$Fe$_{0.23}$Mn$_{1.52}$In$_{0.5}$\\ [1ex]
\hline
\end{tabular}
\label{table:EDX}
\end{table}

\section{Results and Discussion}

\begin{figure}[h]
\begin{center}
\includegraphics[width=\columnwidth]{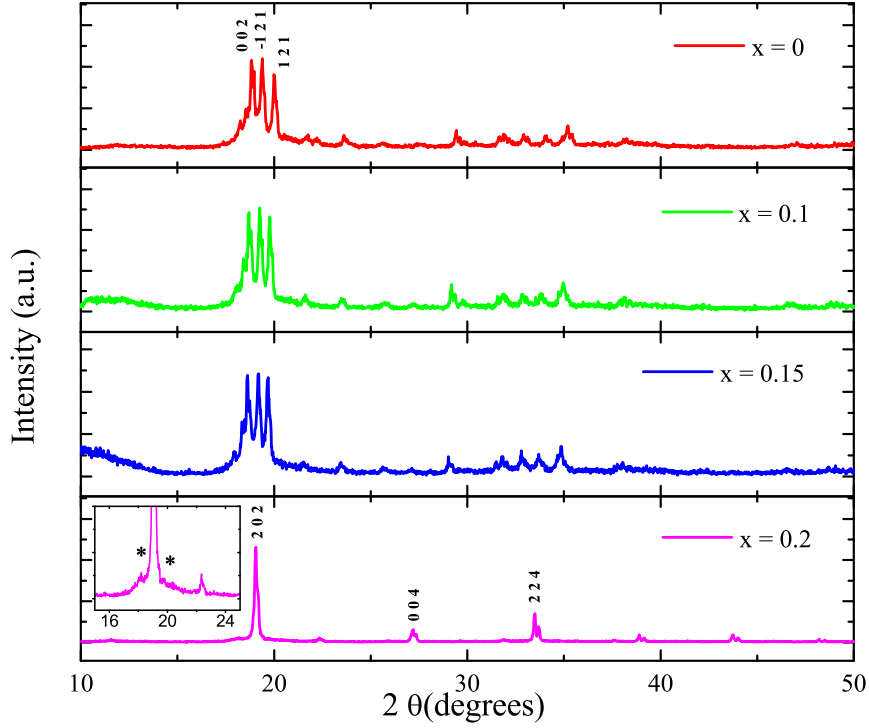}
\caption{X-ray diffraction patterns of Ni$_{2-x}$Mn$_{1.5}$Fe$_x$In$_{0.5}$. Alloys with $x$ = 0, 0.1, 0.15 display incommensurate 7M modulated monoclinic structure and $x$ = 0.2 exhibits cubic structure. Inset shows a presence of a minor impurity phase (marked as $\ast$) in $x$ = 0.2 alloy.}
\label{fig:XRD}
\end{center}
\end{figure}

The x-ray diffraction patterns for all the samples recorded at room temperature are displayed in Fig. \ref{fig:XRD}. Le Bail refinement on these xrd patterns indciate that the samples with $x$ = 0, 0.1 and 0.15 exhibit an incommensurate 7M modulated structure. This implies their martensitic transformation temperature, T$_M$ is above room temperature \cite{Devi201897,Asli2015140} (see Fig. \ref{fig:XRD}a to Fig. \ref{fig:XRD}c). The alloy composition with $x$ = 0.2, on the other hand, exhibits a cubic phase (Fig. \ref{fig:XRD}d) implying its martensitic temperature to be either below room temperature or completely suppressed. The lattice constant of the cubic phase was obtained to be 5.980368(224)\AA~ which is very close to the limiting value of a martensitic transition to be observed in Ni-Mn based Heusler alloys \cite{Planes200921}. Hence a cubic structure at room temperature could be due to complete suppression of the martensitic phase in the alloy with $x = 0.2$. It must be pointed out, that a few additional peaks of very weak intensity (see Fig. \ref{fig:XRD}d) are seen on the two sides of the main peak belonging to a minority phase.

\begin{figure}[h]
\begin{center}
\includegraphics[width=\columnwidth]{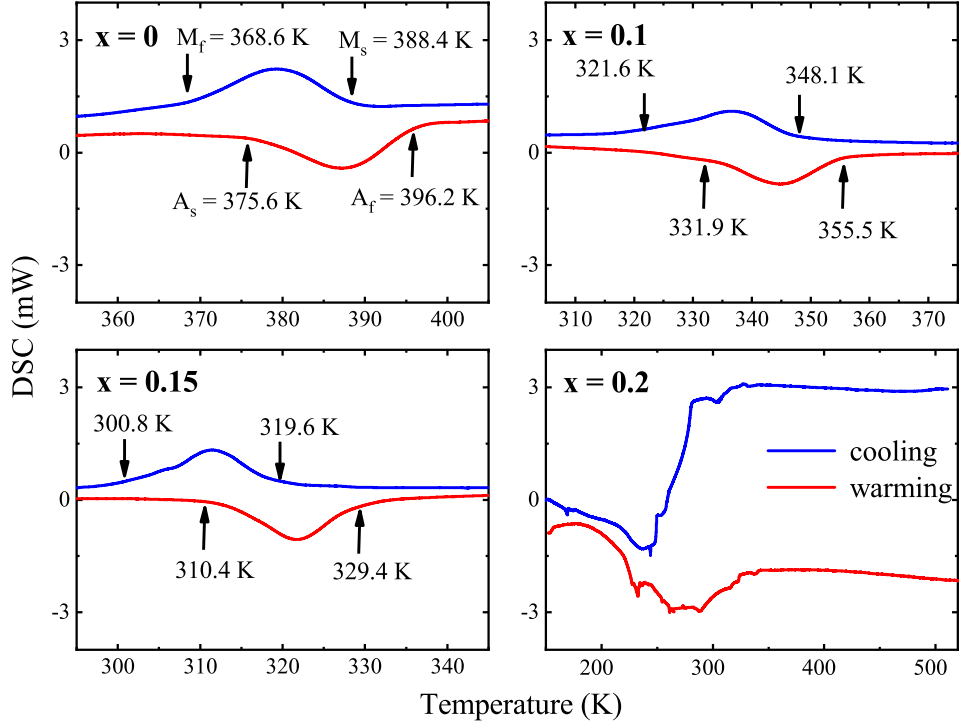}
\caption{Differential scanning calorimetry plots during warming and cooling cycles performed at the rate of 5 degrees/min in Ni$_{2-x}$Mn$_{1.5}$Fe$_x$In$_{0.5}$ ($x$ = 0, 0.1, 0.15, 0.2).}
\label{fig:DSC}
\end{center}
\end{figure}

The austenite to martensite transition temperatures for $x$ = 0, 0.1, 0.15 and 0.2 were concluded through the exothermic and endothermic thermograms of the DSC measurements presented in Fig. \ref{fig:DSC}a to Fig. \ref{fig:DSC}d respectively. The hysteresis in the warming and cooling cycle confirms the first order nature of the transition. The effect of Fe doping is evident from the fact that a slight increase in Fe content results in a sharp decrease in the martensitic transition temperature as seen in Fig. \ref{fig:DSC}. This observation is consistent with earlier results of Fe doping in Ni-Mn based Heusler alloys \cite{Sharma201022, Lobo2014116, Nevgi2018112}. The cubic composition Ni$_{1.8}$Mn$_{1.5}$Fe$_{0.2}$In$_{0.5}$ does not seem to show any transition over the temperature range of 450K-150K as can be seen from Fig. \ref{fig:DSC}d. Instead, a kind of broad feature in the temperature range of 350K to 200K is visible during the warming cycle. Such behavior normally occurs due to the presence of a transition to strain glassy phase\cite{Sarkar200595}. During the cooling cycle, a relatively sharp drop beginning at about 275K is clearly noticed. Though the usual care was taken to avoid any contamination due to moisture, such changes in DSC thermograms could also be due to the presence of a minor amount of moisture content either in the sample environment or in the purge gas.

\begin{figure}[h]
\begin{center}
\includegraphics[width=\columnwidth]{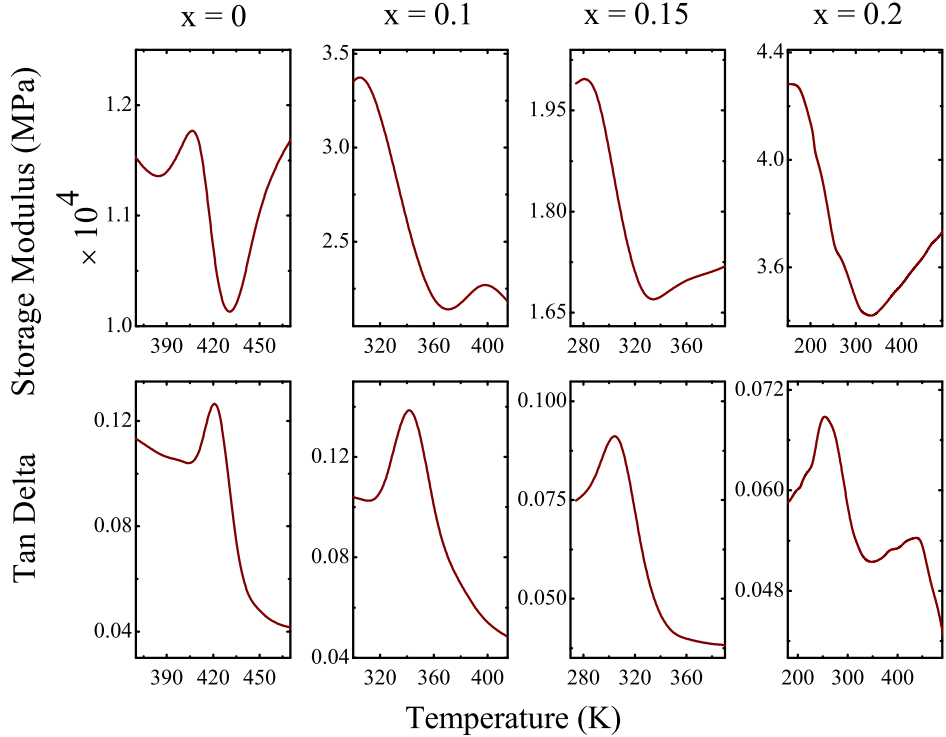}
\caption{Temperature dependent ac storage modulus and tan $\delta$ measurements for the series Ni$_{2-x}$Mn$_{1.5}$Fe$_x$In$_{0.5}$ ($x$ = 0, 0.1, 0.15, 0.2) at a representative frequency of 1.1 Hz.}
\label{fig:DMA}
\end{center}
\end{figure}

\begin{figure}[h]
\begin{center}
\includegraphics[width=\columnwidth]{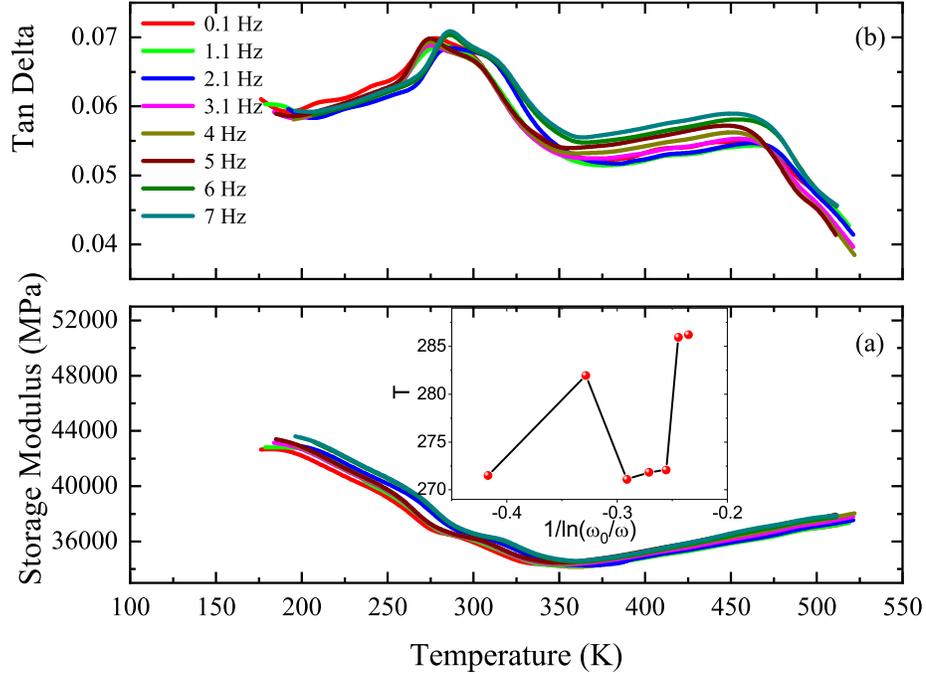}
\caption{Temperature dependent ac storage modulus and tan $\delta$ measurements for the composition Ni$_{1.8}$Mn$_{1.5}$Fe$_{0.2}$In$_{0.5}$ at multiple frequencies. Inset shows plot of $T_g$ versus $\log \omega$ which is not in accordance with Vogel Fulcher relation for glassy dynamics.}
\label{fig:DMA-MF}
\end{center}
\end{figure}

To explore the possibility of a strain glass transition in Ni$_{1.8}$Fe$_{0.2}$Mn$_{1.5}$In$_{0.5}$, DMA measurements were carried out on alloys with $x$ = 0, 0.1, 0.15 and 0.2 at multiple frequencies in the range 0.1 Hz to 7 Hz. The elastic properties, storage modulus and loss ($\tan \delta$) of these samples as a function of temperature at a single frequency of 1.1 Hz are presented in Fig. \ref{fig:DMA}. A sharp dip in the storage modulus and a peak in the loss is seen at the onset of martensitic transition for all the transforming compositions up to $x$ = 0.15. However, the composition $x$ = 0.2  shows a broad valley in the storage modulus and a broad peak in the loss at about 280K. Another broad feature in the loss is seen at 450K, but there is no corresponding feature seen in the storage modulus. The frequency dependence of storage modulus and loss feature at 280K were probed and are displayed in Fig. \ref{fig:DMA-MF}. No Vogel-Fulcher dependence is seen as a function of frequency indicating the absence of a strain glassy transition in this alloy (see inset of Fig. \ref{fig:DMA-MF}). The loss feature at 450K also did not show any frequency dependence in accordance with Vogel-Fulcher law.

\begin{figure}[h]
\begin{center}
\includegraphics[width=\columnwidth]{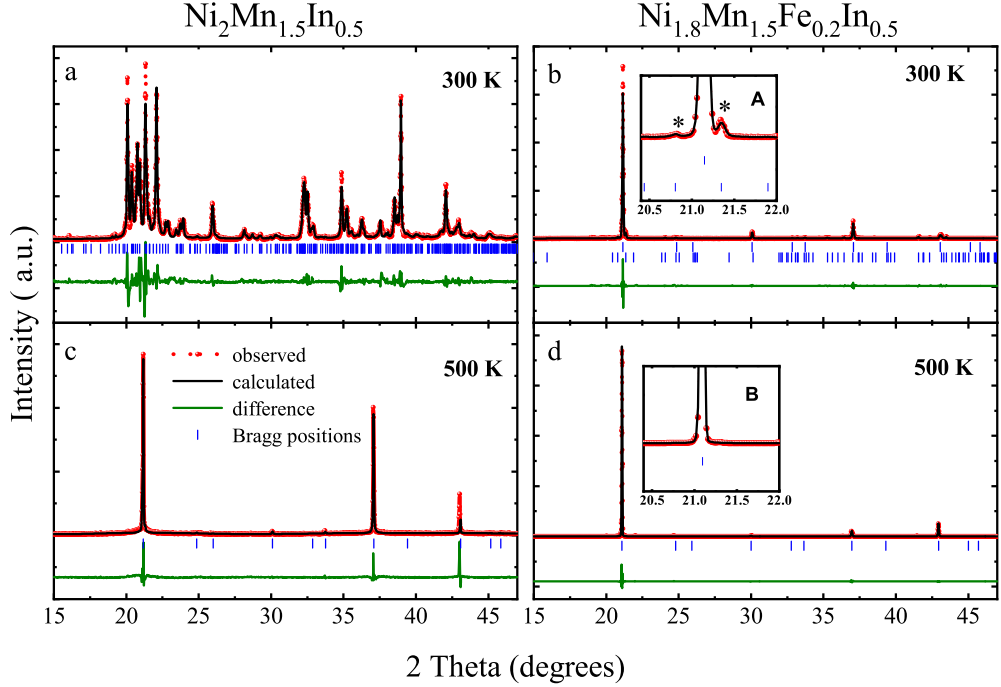}
\caption{The lebail refined data for the compositions ($x$ = 0 and 0.2) in Ni$_{2-x}$Mn$_{1.5}$Fe$_x$In$_{0.5}$ at two temperatures 300 K and 500 K. Inset A shows biphasic composition $x$ = 0.2 at 300K in wherein the secondary phase is marked with $\ast$ which disappears at high temperature (inset B).}
\label{fig:SYXRD}
\end{center}
\end{figure}

Thus Ni$_{1.8}$Fe$_{0.2}$Mn$_{1.5}$In$_{0.5}$ neither displays a martensitic transition nor a strain glassy phase. Further, the possibility of crystallization of strain glassy phase was checked by performing repeated DSC measurements on $x$ = 0.2 alloy after giving it an isothermal treatment at certain temperatures below 300K \cite{Yuanchao1142015}, but the DSC thermograms did not show any significant changes. Hence SEM-EDX measurements were re-looked. The surface elemental mapping revealed the existence of two regions, a minor region severely deficient in in Fe (Ni$_{1.92}$Fe$_{0.08}$Mn$_{1.5}$In$_{0.5}$) embedded in a relatively Fe rich (Ni$_{1.71}$Fe$_{0.29}$Mn$_{1.5}$In$_{0.5}$) matrix. As mentioned earlier, the room temperature xrd performed on laboratory source showed a major cubic phase with a few extra peaks belonging to a minority phase. As a result, a careful Lebail analysis is carried out on the synchrotron xrd data recorded at 300K and 500K and is presented in Fig. \ref{fig:SYXRD}.

At room temperature, while the undoped alloy, Ni$_2$Mn$_{1.5}$In$_{0.5}$ exhibits the expected 7M incommensurate martensitic structure (Fig. \ref{fig:SYXRD}a), Ni$_{1.8}$Fe$_{0.2}$Mn$_{1.5}$In$_{0.5}$ exhibits a coexistence of two phases, a major cubic phase and a minor phase fitted to 7M monoclinic structure in a ratio 80:20 (Fig. \ref{fig:SYXRD}b). The transformation temperature of the undoped alloy being close to 400K, its high temperature (500K) structure is expectedly cubic as can be seen in Fig. \ref{fig:SYXRD}c. Interestingly, the $x = 0.2$ alloy also exhibits a single phase cubic structure at 500K (Fig. \ref{fig:SYXRD}d). A comparison of the two insets, A and B in Fig. \ref{fig:SYXRD}, showing an expanded view of the region near the main (220) reflection of the cubic phase reiterates vanishing of the minority monoclinic phase in Ni$_{1.8}$Fe$_{0.2}$Mn$_{1.5}$In$_{0.5}$ at 500K. This implies that the monoclinic phase fraction transforms to cubic phase at some temperature below 500K. DSC measurements reported in Fig. \ref{fig:DSC} show that all compositions with Fe content, $x < 0.2$ undergo martensitic transition in the temperature interval of 300K to 400K. Therefore the Fe poor phase detected in Ni$_{1.8}$Fe$_{0.2}$Mn$_{1.5}$In$_{0.5}$ at 300K undergoing a transformation to cubic phase in the temperature range from 300K to 500K is not surprising. There is also no significant difference in the lattice constants of the cubic phase of $x$ = 0 and $x$ = 0.2 alloys so only a single cubic phase appears at T $\ge$ 500K. Nucleation of such minor phases has been reported in other alloy families as well \cite{Lu2019}. A structural phase transition of a minor phase embedded in a major non transforming matrix could be the cause of broad transition seen in the DSC thermogram around 300K. The broad nature of the transition could arise due to minor compositional variation of the structural compositions from one region to another. Further support to this argument is gained from the broad feature seen around 350K in the ac storage modulus reported in Fig. \ref{fig:DMA-MF}(a).

\begin{figure}[h]
\begin{center}
\includegraphics[width=\columnwidth]{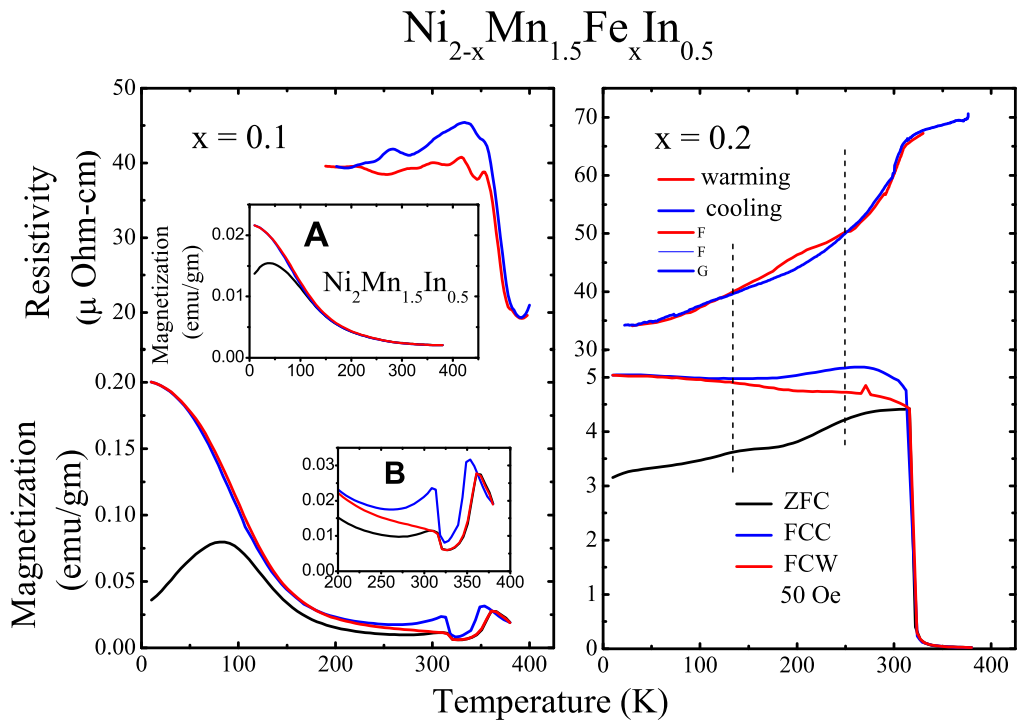}
\caption{Resistivity and magnetization plots for the compositions ($x$ = 0.1, 0.2) in Ni$_{2-x}$Mn$_{1.5}$Fe$_x$In$_{0.5}$. Resistivity data were recorded during warming (red) and cooling (blue) cycles while magnetization data was recorded in 5 mT applied filed during warming after cooling in zero field (ZFC - black) and subsequent cooling (FCC - blue) and warming (FCW - red) cycles. Inset A presents magnetization for the $x$ = 0 composition while Inset B highlights the sharp rise in magnetization in the composition $x$ = 0.1 at 314K.}
\label{fig:RES+MAG}
\end{center}
\end{figure}

Temperature dependent xrd measurements reveal presence of a major Fe rich cubic phase ($\sim$ 80\%) and minor Fe poor martensitic phase ($\sim$ 20\%) in Ni$_{1.8}$Fe$_{0.2}$Mn$_{1.5}$In$_{0.5}$. This is exactly opposite to the situation in strain glassy composition, Ni$_{2}$Mn$_{1.4}$Fe$_{0.1}$In$_{0.5}$. In order to understand the effect of such a reversal of phase fraction on transport and magnetic properties, temperature dependent resistivity and magnetization are measured and presented in Fig. \ref{fig:RES+MAG}. The resistivity of Ni$_2$Mn$_{1+x}$In$_{1-x}$ martensitic alloys displays a sharp jump at the martensitic transition temperature with its magnitude almost doubling below the transition and pronounced hysteresis between the warming and cooling cycles. Those compositions are not undergoing martensitic transition display a typical metallic behavior in the entire temperature range. For a strain glass, a broad anomaly at $T_g$, instead of a sharp increase in resistivity is noted, and it is also devoid of any hysteretic behavior during the warming and cooling cycles. The resistivity plot for the composition with $x$ = 0.1 is presented in Fig. \ref{fig:RES+MAG}a. A sudden increase in resistivity can be seen at the martensitic transition temperature. Hysteresis in resistivity measured during the warming and cooling cycles is also present.  However, in the case of $x$ = 0.2, the resistivity behavior completely changes (Fig. \ref{fig:RES+MAG}b). It displays a metallic behavior with an anomaly at 314K followed by a weak hysteresis that is more prominently seen in the temperature range of 250K to 140K. Such a resistivity behavior could be ascribed to major cubic phase seen from the diffraction measurements (see Fig. \ref{fig:SYXRD}c).

Magnetization as a function of temperature in an applied field of 5 mT during ZFC, FCC and FCW cycles for $x$ = 0.1, and 0.2 compositions are presented in Fig. \ref{fig:RES+MAG} c and d respectively. The temperature dependent magnetization behavior of $x$ = 0.1 is quite similar to that of  the undoped composition ($x$ = 0)  shown in inset A in Fig. \ref{fig:RES+MAG}.  While a clear hysteresis in cooling (FCC) and warming (FCW) cycles around 350K corresponding to the martensitic transition of the alloy can be seen in $x$ = 0.1, no such transition is seen in $x$ = 0. This is presumably because of the high temperature limit of the magnetometer ($\sim$ 380K) is below the martensitic transition temperature of Ni$_2$Mn$_{1.5}$In$_{0.5}$ alloy ($T_M$ = 400 K). On the other hand, the magnetization of Ni$_{1.8}$Fe$_{0.2}$Mn$_{1.5}$In$_{0.5}$, shows a sharp increase corresponding to a paramagnetic to ferromagnetic transition at 314K matching well with the change in slope seen in the resistivity measurement. A small but sharp increase in magnetization at about the same temperature is also seen in $x$ = 0.1 (see inset B of Fig. \ref{fig:RES+MAG}). Fe doping in Ni-Mn based martensitic alloys is shown to strengthen the ferromagnetic austenitic phase and suppress the martensitic transition \cite{Sharma201022, Lobo2014116}. Therefore the ferromagnetic transition in $x$ = 0.2 alloy arises from the magnetic ordering of the majority cubic phase and the presence of hysteresis in the FCC and FCW curves seen in the temperature interval of 350K - 150K could be due to the presence of minor martensitic phase as seen from the synchrotron xrd studies (see Fig. \ref{fig:SYXRD}). The presence of a similar transition at 314K in $x$ = 0.1 could be an indication of similar phase separation as in $x$ = 0.2 alloy. In $x$ = 0.1 the ferromagnetic phase could be a minor phase whose fraction increases with increase of Fe content.

In Ni$_2$Mn$_{1.5}$In$_{0.5}$, strain glassy phase is observed when Fe is doped for Mn resulting in Ni$_{2}$Mn$_{1.5-x}$Fe$_{x}$In$_{0.5}$ while it is absent when Fe is doped for Ni resulting in Ni$_{2-x}$Fe$_{x}$Mn$_{1.5}$In$_{0.5}$. Ni$_{2}$Mn$_{1.4}$Fe$_{0.1}$In$_{0.5}$ though has a modulated monoclinic structure, exhibits glassy dynamics due to the presence of a minor impurity phase, rich in Fe. The impurity phase restricts the long-range interactions between the elastic strain vector resulting in glassy dynamics \cite{Nevgi2018112}. The situation in Ni$_{1.8}$Fe$_{0.2}$Mn$_{1.5}$In$_{0.5}$ is reversed. Synchrotron xrd studies on this alloy indicate cubic phase to be the major phase with only a small fraction ($\sim$ 20\%) exhibiting modulated structure at 300K. Therefore it appears, site symmetry of the dopant atom plays a key role in determining the ground state of such Heusler alloys. When Fe is doped at the X site which has a $\bar43m$ site symmetry, ferromagnetic cubic interactions are promoted leading to a major fraction of the sample remaining in cubic phase. Presence of ferromagnetic interactions suppress martensitic interactions and therefore strain glassy phase is not observed. Whereas, when Fe atoms occupy Y or Z sites in Heusler lattice, ferromagnetic interactions are not promoted and instead result in a minor impurity phase that limits the long range interaction between elastic strain vector leading to a strain glassy phase.

\section{Conclusions}
In conclusion, though impurity doping of Fe at Ni site in Ni$_2$Mn$_{1.5}$In$_{0.5}$ suppresses martensitic transition, the dynamical mechanical properties, in particular, the peak in loss factor does not follow Vogel-Fulcher law as expected for a strain glassy phase. Instead, a crossover from a ground state dominant in antiferromagnetic interactions to a ground state dominant in ferromagnetic interactions is observed. The ferromagnetic interactions arise due to the formation of a major cubic phase relatively rich in Fe as evidenced by synchrotron xrd and magnetization studies. These ferromagnetic interactions  suppress the glassy dynamics leading to the absence of strain glassy phase in Ni$_{1.8}$Fe$_{0.2}$Mn$_{1.5}$In$_{0.5}$.

\section*{Acknowledgements}
The authors wish to acknowledge the financial assistance from the Science and Engineering Research Board, Govt. of India under the project SB/S2/CMP-0096/2013. Authors thank the Department of Science and Technology, India for the travel support, Saha Institute of Nuclear Physics and Jawaharlal Nehru Centre for Advanced Scientific Research, India for facilitating the experiments at the Indian Beamline, Photon Factory, KEK, Japan.

\bibliographystyle{elsarticle-num}
\bibliography{Ref}

\begin{thebibliography}{10}
\expandafter\ifx\csname url\endcsname\relax
  \def\url#1{\texttt{#1}}\fi
\expandafter\ifx\csname urlprefix\endcsname\relax\def\urlprefix{URL }\fi
\expandafter\ifx\csname href\endcsname\relax
  \def\href#1#2{#2} \def\path#1{#1}\fi

\bibitem{Zhou200995}
Y.~Zhou, D.~Xue, X.~Ding, K.~Otsuka, J.~Sun, X.~Ren, Applied Physics Letters 95
  (2009) 151906 (2009).

\bibitem{Zhou201058}
Y.~Zhou, D.~Xue, X.~Ding, Y.~Wang, J.~Zhang, Z.~Zhang, D.~Wang, K.~Otsuka,
  J.~Sun, X.~Ren, Acta Materialia 58 (2010) 5433--5442 (2010).

\bibitem{Sarkar200595}
S.~Sarkar, X.~Ren, K.~Otsuka, Phys. Rev. Lett. 95 (2005) 205702 (2005).

\bibitem{Lloveras2008100}
P.~Lloveras, T.~Castan, M.~Porta, A.~Planes, A.~Saxena, Phys. Rev. Lett. 100
  (2008) 165707 (2008).

\bibitem{Vasseur201082}
R.~Vasseur, T.~Lookman, S.~R. Shenoy, Phys. Rev. B 82 (2010) 094118 (2010).

\bibitem{Wang201058}
D.~Wang, Z.~Zhang, J.~Zhang, Y.~Zhou, Y.~Wang, X.~Ding, Y.~Wang, X.~Ren, Acta
  Materialia 58 (2010) 6206--6215 (2010).

\bibitem{MA201874}
H.~Ma, J.~Yang, F.~Lu, F.~Qin, W.~Xiao, X.~Zhao, Progress in Natural Science:
  Materials International 28 (2018) 74 -- 77 (2018).

\bibitem{Ren2012}
X.~Ren, Springer Berlin Heidelberg.

\bibitem{Wang200867}
Y.~Wang, X.~Ren, K.~Otsuka, Materials Science Forum 583 (2008) 67 (2008).

\bibitem{Ren201090}
X.~Ren, Y.~Wang, Y.~Zhou, Z.~Zhang, D.~Wang, G.~Fan, K.~Otsuka, T.~Suzuki,
  Y.~Ji, J.~Zhang, Y.~Tian, S.~Hou, X.~Ding, Philosophical Magazine 90 (2010)
  141--157 (2010).

\bibitem{Nevgi2018112}
R.~Nevgi, K.~R. Priolkar, Applied Physics Letters 112~(2) (2018) 022409 (2018).

\bibitem{Wang201298}
D.~P. Wang, X.~Chen, Z.~H. Nie, N.~Li, Z.~L. Wang, Y.~Ren, Y.~D. Wang,
  Europhysics Letters 98 (2012) 46004 (2012).

\bibitem{Devi201897}
P.~Devi, S.~Singh, B.~Datta, S.~W. D'Souza, Y.~Ikeda, E.~Saurd, V.~Petricek,
  P.~Simon, P.~Werner, S.~Chadhov, S.~Parkin, C.~Felser, D.~Pandey, Phys. Rev.
  B 92 (2018) 224102 (2018).

\bibitem{Asli2015140}
A.~\c{C}akir, L.~Righi, F.~Albertini, M.~Acet, M.~Farle, Acta Materialia 99
  (2015) 140--149 (2015).

\bibitem{Planes200921}
A.~Planes, L.~Mañosa, M.~Acet, Journal of Physics: Condensed Matter 21~(23)
  (2009) 233201 (2009).

\bibitem{Sharma201022}
V.~K. Sharma, M.~K. Chattopadhyay, S.~K. Nath, K.~J.~S. Sokhey, R.~Kumar,
  P.~Tiwari, S.~B. Roy, Journal of Physics: Condensed Matter 22~(48) (2010)
  486007 (2010).

\bibitem{Lobo2014116}
D.~N. Lobo, K.~R. Priolkar, S.~Emura, A.~K. Nigam, Journal of Applied Physics
  116 (2014).

\bibitem{Yuanchao1142015}
J.~Yuanchao, D.~Wang, X.~Ding, K.~Otsuka, X.~Ren, Phys. Rev. Lett. 114 (2015)
  055701 (2015).

\bibitem{Lu2019}
P.~L{\"u}, H.~P. Wang, B.~Wei, Metallurgical and Materials Transactions A
  50~(2) (2019) 789--803 (Feb 2019).

\end{thebibliography}

\end{document}